# The Bi-Spinor Standard Model with 3 Generations

A. Jourjine[1]

FG CTP
Hofmann Str. 6-8
01281 Dresden
Germany

**Abstract**

We show that if two of four generations of the bi-spinor Standard Model are mass-degenerate or sufficiently close in mass only three generations can be observed. We argue that the Standard Model and its bi-spinor analog are indistinguishable on the level of the electroweak precision variables *S*, *T*, *U*. As a result the bi-spinor Standard Model, which describes experimental observed textures of flavor mixing matrices is a better fit to the data then the Standard Model, where the textures are arbitrary.

## 1. Introduction

The origin of the quark and lepton mixing matrices and the difference in their textures has been a long-standing puzzle in elementary particle physics [1, 2, 3, 4]. Its resolution remains elusive and it is one of the main mysteries of the Standard Model (SM). Attempts to explain flavor mixing usually employ extra degrees of freedom. For example, by an exhaustive search of all reasonable discrete symmetry groups it is possible to find a good fit with the experimentally observed data on flavor mixing [5]. However, the many parameters that are brought along make such fits less satisfactory.

Recently, a new approach has been proposed to solve the flavor mixing puzzle, where mixing arises from a hidden conformal symmetry [6]. The approach is based on the use of the bi-spinor Standard Model (b-SM), where the fundamental fermionic fields are described by anti-commuting differential forms [7]. It has been shown that if the fourth generation can be disregarded one obtains a reasonable approximation to the flavor mixing matrices both for quarks and for leptons at the same time. The difference between lepton and quark mixing is explained by the assignment of different value of a new quantum number, called the scalar spin, to lepton and quark generations. Both in the SM and the b-SM the elementary fermions acquire mass after Higgs field transition to a stable vacuum. However, while in the SM the arising complex mass matrices are arbitrary, in the b-SM they are constrained; they must belong to the unitary group $U(2,2)$, which is up to $U(1)$ factor is the conformal group: $SU(2,2)\approx SO(2,4)$.

The bi-spinor approach has two drawbacks. The first is that the parameter space for the forth generation is rather small, if one assumes that the fourth generation has not been observed because of it large mass. This is all but ruled out for the Standard Model and only a few corners in the parameter space are available in its 2HDM extension.

[1] jourjine@pks.mpg.de

The second drawback is the unusual structure of the Lagrangian of the b-SM. One obtains the Lagrangian for the four generations b-SM by subtracting from the sum of the SM-identical fermionic Lagrangians for the first two generations the same for the second two fermionic generations. As a result, b-SM has a hidden local and an obvious global non-compact $U(2,2)$ symmetry. Such theories are usually disregarded in general and for good physical reasons. However, in the case of b-SM the local $U(2,2)$ disappears after extraction of the Dirac degrees of freedom, while the global $U(2,2)$ symmetry is symmetry only on the classical level, application of symmetry transformations leads to inequivalent representations in Fock space: one has to change mass of the particle to effect the transformation. Here the same physical effect is observed with in the Lorentz symmetry. While all frames of reference result in the same Lagrangian, but to physically go from one frame of reference to another one requires an acceleration and acceleration is gravitational in nature and lies outside of the scope of the standard Minkowski field theory.

As a result of the imbedded $U(2,2)$ b-SM is not an extension of the SM. The theory contains two generations that propagate forward in time and two generations that propagate backward in time. The field content is reminiscent of the doubling of the degrees of freedom in Keldysh [8] treatment of non-equilibrium quantum field theory or in its more formal analog, the thermofield theory [9].

In this letter we propose solutions for both problems. We show that if two of the four generations have equal mass, then they become a pair of particle and antiparticle. Therefore, in such a case the fourth generation is in effect unobservable. The same occurs when two masses are sufficiently close in value.

Secondly, we argue that there are no presently detectable differences that appear when one generation enters in the Lagrangian with a minus sign: for the b-SM that would be the third generation of leptons and the third generation of quarks. The most important of these is the influence of such a change on the precision electroweak variables, *S, T, U* [10]. We point out that since b-SM is not an extension of the SM, the use of *S, T, U* is unrevealing, for they can only measure distances in the space of all SM extensions.

What remains of the changes in loop diagrams is that the propagators for heaviest flavors are modified by change of the sign and by change in the contour of integration in the momentum space: $i\varepsilon \to -i\varepsilon$. This reflects the fact that the minus generation propagates backwards in time. Note that these two changes do not affect the 1-loop Feynman integrals for electroweak boson self-energies used in the definition of the EW precision variables. Therefore, the experimentally observable differences with the SM appear at a higher loop level. They can only come when the forward-time particles of the first two generations interfere with the backward-time particles of the third generation in the sums that enter loop amplitude computations. Therefore, we conclude that on the level of EW precision variables b-SM and SM are indistinguishable.

Let us now turn to the question of the fourth generation. We will now show that under certain conditions the extra-generation behaves like the second generation for all intends and purposes. To start with we recapitulate the results in [6, 7].

After spontaneous symmetry breakdown the free-field fermionic part of the b-SM is

$$\mathcal{L}_{b-SM} = \tilde{\mathcal{L}}_q + \tilde{\mathcal{L}}_l , \tag{1}$$

$$\tilde{\mathcal{L}}_q = \overline{\overline{Q}}_i^A (i\partial) Q_i^A + \overline{\overline{u}}_R^A (i\partial) u_R^A + \overline{\overline{d}}_R^A (i\partial) d_R^A - \left( \overline{\overline{Q}}_1^A \mathrm{M}_u^{AB} u_R^B + \overline{\overline{Q}}_2^A \mathrm{M}_d^{AB} d_R^B + c.c. \right), \tag{2}$$

$$\tilde{\mathcal{L}}_l = \overline{\overline{E}}_i^A (i\partial) E_i^A + \overline{\overline{\nu}}_R^A (i\partial) \nu_R^A + \overline{\overline{e}}_R^A (i\partial) e_R^A - \left( \overline{\overline{E}}_1^A \mathrm{M}_\nu^{AB} \nu_R^B + \overline{\overline{E}}_2^A \mathrm{M}_e^{AB} e_R^B + c.c. \right), \tag{3}$$

where $\overline{\overline{Q}}^{A} = \Gamma^{AB}\overline{Q}^{B}$, and $\Gamma^{AB} = diag(1, 0)$ for Dirac spinors, $\Gamma^{AB} = diag(0, -1)$ for anti-Dirac spinors, and $\Gamma^{AB} \equiv {\sigma_3}^{AB} = diag(1, -1)$ for a Dirac-anti-Dirac doublet spinors. Dirac and anti-Dirac spinors describe a single generation each, while Dirac-anti-Dirac doublet describes two generations, but a single elementary fermion. Any other free-field fermionic Lagrangian can be formed by a adding together arbitrary number of generations the three types above. We refer to [] for the origin of the terminology and explanations.

As a consequence of $U(2,2)$ symmetry, in the b-SM the mass matrices $\mathrm{M}_{u,d}$, $\mathrm{M}_{l,\nu}$ are not arbitrary but have a specific form

$$\mathrm{M} = m B_1 \mathcal{M} B_2 . \tag{4}$$

where $m$ is a parameter with dimension of mass, $B_a \in U(2) \oplus U(2)$, $a = 1, 2$ are $4 \times 4$ block-diagonal matrices with the upper-left blocks of which mix only $A = 1, 2$, while the lower-right blocks mix only $A = 3, 4$. Factors $B_a$ are arbitrary. They have the same form for both quark and lepton sectors

$$B_a = \begin{pmatrix} U_1^a & 0 \\ 0 & U_2^a \end{pmatrix}, \qquad U_k^a = \begin{pmatrix} {x_k}^a & {y_k}^a \\ {z_k}^a & {w_k}^a \end{pmatrix} \in U(2), \qquad a, k = 1,2 . \tag{5}$$

Dimensionless matrix $\mathcal{M}$ is also a direct sum of two two-dimensional matrices but now the first summand mixes generations 1 and 3 only, while the second summand mixes generations 2 and 4

$$m\,\mathcal{M} = m_1\, {\mathcal{M}_R}^{(p)} \oplus m_2\, {\mathcal{M}_R}^{(q)}, \qquad p, q = 1,2 , \tag{6}$$

where ${\mathcal{M}_R}^{(1)}$ is diagonal and ${\mathcal{M}_R}^{(2)} \in U(1,1)$

$${\mathcal{M}_R}^{(2)} = \begin{pmatrix} c_\lambda & s_\lambda \\ s_\lambda & c_\lambda \end{pmatrix}, \qquad s_\lambda = \sinh\lambda, \quad c_\lambda = \cosh\lambda, \quad {c_\lambda}^2 - {s_\lambda}^2 = 1 . \tag{7}$$

Requiring that fermions have physical masses results in that there are only four possible cases for $\mathcal{M}$ given by

$$\begin{aligned} &m\,\mathcal{M} = m_1\, {\mathcal{M}_R}^{(1)} \oplus m_2\, {\mathcal{M}_R}^{(1)\prime}, \qquad && m\,\mathcal{M} = m_1\, {\mathcal{M}_R}^{(1)} \oplus m_2\, {\mathcal{M}_R}^{(2)} \\ &m\,\mathcal{M} = m_2\, {\mathcal{M}_R}^{(2)} \oplus m_1\, {\mathcal{M}_R}^{(1)}, \qquad && m\,\mathcal{M} = m_1\, {\mathcal{M}_R}^{(2)} \oplus m_2\, {\mathcal{M}_R}^{(2)\prime} , \end{aligned} \tag{8}$$

where prime denotes a matrix with different non-zero entries. The possible mass matrices are listed in order of increasing mass degeneracy. The first case has 4 independent mass parameters, the second and the third three mass parameters, while the fourth has two mass parameters. In the limiting case $m_1 = m_2 \quad \mathcal{M} \subset U(2,2)$.

Mixing matrices for quarks and leptons are defined in the b-SM by

$$V = U_L D_L^+, \qquad U = E_L N_L^+ . \tag{9}$$

In the SM transformation to mass basis is defined as a transformation that decouples the fields of different generations in the Lagrangian. After the transformation the free field fermionic SM Lagrangian becomes a sum of four Lagrangians each containing the fields for one of the four generations. In bi-spinor theory the situation is somewhat different. While for diagonal mass matrix summand $\mathcal{M}_R^{(1)}$ in (6) the diagonalization means exactly the same as in SM, for the $\mathcal{M}_R^{(2)}$ in (6) such diagonalization is impossible, because the corresponding kinetic term bilinear matrix $\sigma_3 = diag(1,-1)$ does not commute with all possible unitary transformations that diagonalize mass matrices.

Therefore, the definition of diagonalization is modified. It is defined as diagonalization of the equations of motion. This definition is sufficient for definition of mass eigenstates.

The transformation that decouples equations of motion for $\mathcal{M}_R^{(2)}$ case is given by

$$W^{(2)} = \frac{1}{\sqrt{2}}\begin{pmatrix} 1 & -1 \\ 1 & 1 \end{pmatrix}, \tag{10}$$

where $W$ mixes either indexes 1 and 3 or 2 and 4. Note that the order in which stripping of the unitary factors is fixed. First comes stripping of the unitary factors in (5) and then transformation (10). For $\mathcal{M}_R^{(1)}$ case we may use the unit matrix

$$W^{(1)} = \begin{pmatrix} 1 & 0 \\ 0 & 1 \end{pmatrix}. \tag{11}$$

Therefore, the generic diagonalizing transformation that corresponds to the four possible mass matrices is

$$T^{(p,q)(r,s)} = U_L D_L^{+} = \left(W_U{}^{(p,q)}\tilde{U}_L\right)\left(W_D{}^{(r,s)}\tilde{D}_L\right)^{+}, \quad p,q,r,s = 1,2\,, \tag{12}$$

where $\tilde{U}_L, \tilde{D}_L$ are block-diagonal

$$\tilde{U}_L = \begin{pmatrix} U_1 & 0 \\ 0 & U_2 \end{pmatrix} \in U(2)\oplus U(2),$$
$$\tilde{D}_L = \begin{pmatrix} D_1 & 0 \\ 0 & D_2 \end{pmatrix} \in U(2)\oplus U(2), \quad U_k, D_k \in U(2). \tag{13}$$

A convenient expression for $W^{(p,q)}$ is given if we swap generation 2 and 3. Then instead of mixing generations 1 and 3 or 2 and 4 matrices $\mathcal{M}_R^{(k)}$ mix generations 1 and 2 or 3 and 4. After such generation swap $W^{(p,q)}$ becomes block-diagonal

$$W^{(p,q)} = \begin{pmatrix} \mathcal{M}_R{}^{(p)} & 0 \\ 0 & \mathcal{M}_R{}^{(q)} \end{pmatrix}. \tag{14}$$

Since $W \equiv \tilde{U}_L \tilde{D}_L{}^{+} \in U(2) \oplus U(2)$ is arbitrary we can write down $T_Q$, the generic quark, or $T_L$, the generic lepton mixing matrix, as

$$T_{Q,L}{}^{(p,q)(r,s)} = W^{(p,q)} W_{Q,L} \left(W^{(r,s)}\right)^{+}, \quad p,q,r,s = 1,2, \quad W_{Q,L} = \begin{pmatrix} U_1^{Q,L} & 0 \\ 0 & U_2^{Q,L} \end{pmatrix}, \tag{15}$$

where the upper-left block of $W_{Q,L}$ mixes generations 1 and 2, while its lower-right block mixes generations 3 and 4. Matrix $W^{(p,q)}$ also satisfies $W^{(p,q)} \in U(2) \oplus U(2)$. However, in (14) the first block in $W^{(p,q)}$ mixes generations 1 and 3, while the second generations 2 and 4. Explicitly, the four possible matrices $W^{(p,q)}$ are given by

$$W^{(0,0)} = \begin{pmatrix} 1 & 0 & 0 & 0 \\ 0 & 1 & 0 & 0 \\ 0 & 0 & 1 & 0 \\ 0 & 0 & 0 & 1 \end{pmatrix}, \qquad W^{(0,1)} = \begin{pmatrix} 1/\sqrt{2} & 0 & -1/\sqrt{2} & 0 \\ 0 & 1 & 0 & 0 \\ 1/\sqrt{2} & 0 & 1/\sqrt{2} & 0 \\ 0 & 0 & 0 & 1 \end{pmatrix}, \tag{16}$$

$$W^{(1,0)} = \begin{pmatrix} 1 & 0 & 0 & 0 \\ 0 & 1/\sqrt{2} & 0 & -1/\sqrt{2} \\ 0 & 0 & 1 & 0 \\ 0 & 1/\sqrt{2} & 0 & 1/\sqrt{2} \end{pmatrix}, \qquad W^{(1,1)} = \begin{pmatrix} 1/\sqrt{2} & 0 & -1/\sqrt{2} & 0 \\ 0 & 1/\sqrt{2} & 0 & -1/\sqrt{2} \\ 1/\sqrt{2} & 0 & 1/\sqrt{2} & 0 \\ 0 & 1/\sqrt{2} & 0 & 1/\sqrt{2} \end{pmatrix}.$$

It follows that altogether there are 16 possible types of mixing matrices in 4-generation bi-spinor theory that differ in texture. They are all parameterized by arbitrary block-diagonal $W_{Q,L}$. It is not difficult to find the best fit of the assignment of scalar spin values to leptons and quarks. This produces the four-dimensional mixing matrices

$$\hat{V}_{CKM}^{4} \equiv (1,1)(1,1) = \frac{1}{2} \begin{pmatrix} x_1 + x_2 & y_1 + y_2 & x_1 - x_2 & y_1 - y_2 \\ z_1 + z_2 & w_1 + w_2 & z_1 - z_2 & w_1 - w_2 \\ x_1 - x_2 & y_1 - y_2 & x_1 + x_2 & y_1 + y_2 \\ z_1 - z_2 & w_1 - w_2 & z_1 + z_2 & w_1 + w_2 \end{pmatrix}, \tag{17}$$

$$\hat{U}_{PMNS}^{4} \equiv (1,0)(0,0) = \begin{pmatrix} \hat{x}_1 & \hat{y}_1 & 0 & 0 \\ \hat{z}_1/\sqrt{2} & \hat{w}_1/\sqrt{2} & -\hat{z}_2/\sqrt{2} & -\hat{w}_2/\sqrt{2} \\ 0 & 0 & \hat{x}_2 & \hat{y}_2 \\ -\hat{z}_1/\sqrt{2} & -\hat{w}_1/\sqrt{2} & -\hat{z}_2/\sqrt{2} & -\hat{w}_2/\sqrt{2} \end{pmatrix}, \tag{18}$$

We now note that if masses of two out of four generations coincide the fourth generation becomes an anti-particle for its partner in the doublet. Therefore, the fourth generation will not be observed as an extra generation but as its doublet partner: the number of generations is thus effectively reduced from four to three. It is reasonable to expect that the pattern of flavor mixing will persist when the masses of two generations are no exactly equal but differ in value by a sufficiently small amount. Therefore, we arrive at the final form of $3\times3$ mixing matrices for the b-SM with 4 or effective 3 generations

$$V_{b-CKM} = \frac{1}{2}\begin{pmatrix} x_1 + x_2 & y_1 + y_2 & y_1 - y_2 \\ z_1 + z_2 & w_1 + w_2 & w_1 - w_2 \\ z_1 - z_2 & w_1 - w_2 & w_1 + w_2 \end{pmatrix}, \tag{19}$$

$$U_{b-PMNS} = \begin{pmatrix} \hat{x}_1 & \hat{y}_1 & 0 \\ \hat{z}_1/\sqrt{2} & \hat{w}_1/\sqrt{2} & -\hat{w}_2/\sqrt{2} \\ -\hat{z}_1/\sqrt{2} & -\hat{w}_1/\sqrt{2} & -\hat{w}_2/\sqrt{2} \end{pmatrix}. \tag{20}$$

For discussion of how well these matrices describe the experimental values we refer the reader to [6].